\documentclass[aps,pra,twocolumn, 10pt, showpacs,superscriptaddress,groupedaddress]{revtex4-1}

\newcommand{\cm}{\mbox{ cm}}
\newcommand{\mm}{\mbox{ mm}}

\newcommand{\nm}{\mbox{ nm}}
\newcommand{\ms}{\mbox{ ms}}

\newcommand{\Hz}{\mbox{ Hz}}

\newcommand{\GHz}{\mbox{ GHz}}

\newcommand{\mW}{\mbox{ mW}}

\newcommand{\nK}{\mbox{ nK}}

\newcommand{\Rb}{\ensuremath{^{87}}Rb }
\newcommand{\tp}{\ensuremath{\tau}_p }
\newcommand{\pes}{\sigma_P^\text{es}}
\newcommand{\pth}{\sigma_P^\text{th}}
\newcommand{\ket}[1]{\ensuremath{\left| #1 \right>}}

\usepackage{graphicx}
\usepackage{epstopdf}
\usepackage{bbold}
\usepackage{comment}
\usepackage{color}
\usepackage{soul}
\usepackage[abs]{overpic}
\usepackage{amsmath,amsfonts,amssymb}
\usepackage{lipsum}
\usepackage{hyperref}
\usepackage{setspace}
\graphicspath{{Figures/}}
\begin{document}
\author{Noam Matzliah}
\affiliation{Department of Physics of Complex Systems, Weizmann Institute of Science, Rehovot 7610001 , Israel }
\author{Hagai Edri}
\affiliation{Department of Physics of Complex Systems, Weizmann Institute of Science, Rehovot 7610001 , Israel }
\author{Asif Sinay}
\affiliation{Department of Physics of Complex Systems, Weizmann Institute of Science, Rehovot 7610001 , Israel }
\author{Roee Ozeri}
\affiliation{Department of Physics of Complex Systems, Weizmann Institute of Science, Rehovot 7610001 , Israel }
\author{Nir Davidson}
\affiliation{Department of Physics of Complex Systems, Weizmann Institute of Science, Rehovot 7610001 , Israel }
\title[]{Observation of Optomechanical Strain in a Cold Atomic Cloud}
\pacs{}
\begin{abstract}
We report the observation of optomechanical strain applied to thermal and quantum degenerate \Rb atomic clouds when illuminated by an intense, far detuned
homogeneous laser beam. In this regime the atomic cloud acts as a
lens which focuses the laser beam. As a backaction, the atoms experience a force opposite to the beam deflection, which depends on the atomic cloud density
profile. We experimentally demonstrate the basic features of this force, distinguishing it from the well-established scattering and dipole forces. The observed
strain saturates, ultimately limiting the momentum impulse that can be transferred to the atoms. This optomechanical force may effectively induce interparticle
interactions, which can be optically tuned.
\end{abstract}
\maketitle
Light-matter interactions are at the core of cold atom physics. A laser beam illuminating atoms close to atomic resonance frequency will apply a scattering force
on them, and an inhomogeneous laser beam far from resonance will mainly apply an optical dipole force~\cite{foot2005atomic}. An intense, far detuned homogeneous
laser beam does not exert a significant force on a single atom, though when applied on inhomogeneous atomic clouds, it will. This was pointed
out~\cite{KetterlePCI} while studying lensing by cold atomic clouds in the context of nondestructive imaging.

The atom's electric polarizability makes atomic clouds behave as refractive media with an index locally dependent on the cloud density. An atomic cloud thus
behaves as a lens that can focus or defocus the laser beam. The atoms recoil in the opposite direction to the beam deflection due to momentum conservation. In
solid lenses, this optomechanical force causes a small amount of stress with negligible strain, due to their rigidity. An atomic lens, however, deforms, making
the force on the atoms observable by imaging their strain. We refer to this optomechanical force as electrostriction, since it resembles shape changes of
materials under the application of a static electric field. Electrostriction can be viewed as an optically induced force between atoms, since the force each atom
experiences depends on the local density of the other atoms.

Optomechanical forces are applied in experiments on refractive matter mainly by optical tweezers, pioneered by~\cite{ashkin1970opticalTweezers}, using structured
light. Less commonly, such forces can be applied by homogeneous light using angular momentum conversion due to the material
birefringence~\cite{CalciteRubinszteinDunlop}, or using structured refractive material shapes~\cite{opticalLift}. Optomechanical forces implemented by such
techniques are used for optically translating and rotating small objects. By applying electrostriction on cold atoms we gain access to the effect of optical
strain - an aspect in optomechanics not directly studied yet in spite of its importance in current research~\cite{OptomechCrystals}.

Interactions between cold atoms can appear naturally or be externally induced and tuned. Tuning is mostly done using a magnetic Feshbach resonance, which was used
to demonstrate many important physical effects such as Bose-Einstein condensate (BEC) collapse and explosion~\cite{donley2001dynamics}, Feshbach
molecules~\cite{donley2002atom,regal2003creation}, BEC-BCS crossover in strongly interacting degenerate
fermions~\cite{greiner2003emergence,jin2003measurement,regal2004observation}, and Fermi
superfluidity~\cite{miller2007critical,siegl2014critical,ferrier2014mixture,delehaye2015critical}. Interactions are also tuned by optical Feshbach
resonance~\cite{bauer2009control}, optical cavities~\cite{ritsch2013cold}, or radio frequency Feshbach resonance~\cite{ding2017effective}. Interactions can be
induced by shining a laser beam on the atoms and creating a feedback mechanism to their response by an externally pumped cavity or a half cavity
\cite{nagy2006self,tesio2012spontaneous,tesio2014kinetic,robb2015quantum,labeyrie2014optomechanical,camara2015optical}. The electrostriction force reported here
is a new kind of induced force between atoms, and may be useful in cold atoms and quantum degenerate atom experiments.

In this Letter we analyze and measure for the first time the optomechanical strain induced in a cold atomic cloud by a homogeneous laser beam far detuned from
atomic resonance. We shine the beam on the cloud and directly observe the resulting strain after time of flight by absorption imaging. We show that this is a new
kind of light-induced force acting on cold atoms. A saturation of the strain is observed, which depends only on the ratio between the momentum impulse applied to
the atomic cloud and the initial momentum distribution width of the cloud. Possible implications for this new force are suggested, and, in particular,
light-induced interaction tuning.

With respect to laser light far from resonance, an inhomogeneous atomic cloud behaves as a lens~\cite{KetterlePCI}, as predicted by the optical Bloch equations.
When a plane wave passes through the cloud, it acquires a position-dependent phase $\phi(\vec{r})$. If the phase is small, the Poynting vector direction
changes~\cite{born1980principles} by an angle $|\vec{\nabla}_{\perp}\phi|/k_L$, where $\vec{\nabla}_{\perp}$ is the gradient along the two directions
perpendicular to the laser beam propagation direction, and $k_L$, the wave number of the beam. As a backaction, the atomic momentum changes in the opposite
direction. The momentum change of the atoms is associated with the electrostriction force, which takes the form~\cite{Note1}
\begin{equation}\label{eq. es}
\begin{split}
    &\vec{f}_\text{es}=\frac{\hbar\Gamma^2}{8\Delta}\frac{I}{I_s}\frac{\vec{\nabla}_{\perp}n}{n}=-\vec{\nabla}_{\perp}U_\text{es}\\
    &U_\text{es}=-\frac{\hbar\Gamma^2}{8\Delta}\frac{I}{I_s}\ln\left(\frac{n}{n_\text{0}}\right),
\end{split}
\end{equation}
where $n_\text{0}$ is an arbitrarily chosen constant density that fixes the arbitrariness in defining a potential up to a constant, $\Gamma$, the width of the
atomic transition, $\Delta$, the detuning of the laser, $I$, its intensity, $I_\text{s}$, the \Rb saturation intensity, and $n$, the local density of atoms.

This force acts only in the directions transverse to the beam propagation and is derived from a potential in the transverse directions that scales logarithmically
with the density. It is a collective force in the sense that it acts only on atoms consisting of an inhomogeneous atomic cloud. The laser induces interactions
between the atoms and the resulting force is independent of the number of atoms. The force scales as $I/\Delta$, similar to the dipole force, and unlike other
light-induced interactions predicted before~\cite{mazets2000ground,kurizki2002new,Cattani}, which are second order in atom-light coupling. For convex clouds it is
repulsive for red detuned laser $\Delta<0$, and attractive for blue detuned laser $\Delta>0$, opposite to the dipole force. Similar to the dipole force, changing
the polarization has a small effect of coupling different atomic states, which effectively changes $I_s$.

\begin{figure}
    \begin{center}
        \begin{overpic}
		    [width=\linewidth]{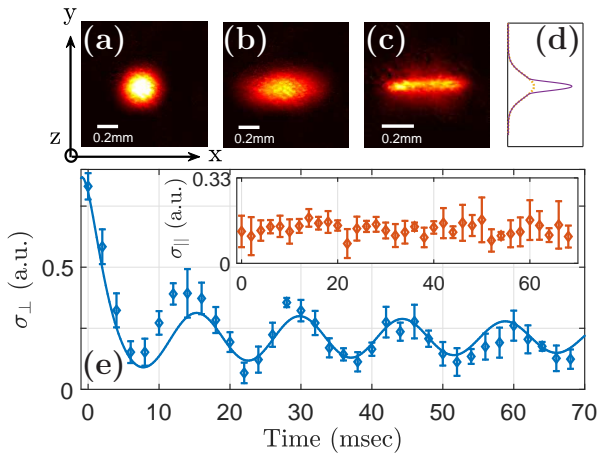}
		    \put(30,156){\large \textcolor{white}{\textbf{(a)}}}
		    \put(83,156){\large \textcolor{white}{\textbf{(b)}}}
		    \put(137,156){\large \textcolor{white}{\textbf{(c)}}}
		    \put(201,156){\large \textbf{(d)}}
		    \put(30,33){\large \textbf{(e)}}
	    \end{overpic}
        \caption{Strain measurements. Absorption image of a thermal cloud after long expansion times with \text{(b)} and without \text{(a)} an electrostriction
        pulse. The cloud aspect ratio changes from unity to $2.1$. We used a laser beam shone along the $\hat{y}$ axis with intensity $8\times10^3\mW/\cm^2$
        ,detuning $47\GHz$ and pulsed for $0.5\ms$. \text{(c)} A BEC after an electrostriction pulse and long expansion time. Even for a strong impulse and large
        aspect ratio the BEC remains partly condensed, showing a bimodal distribution in the axial direction \text{(d)}. \text{(e)} Oscillations in the cloud size
        along one transverse direction (axial direction shown in inset) induced by an electrostriction pulse as a function of a variable waiting time in the trap
        after applying the pulse. A pure transverse breathing mode is observed, fitting to a decaying oscillation (solid line) of twice the trap frequency.}
        \label{Fig. firstfig3}
    \end{center}
\end{figure}

In the experiment we typically trap $10^6$ \Rb atoms in the $\ket{F=1,m_F=1}$ ground state of the $5^2S_{1/2}$ manifold at a temperature of $T=400\nK$. Our
crossed dipole trap has typical trap frequencies of $\omega_x=\omega_y=2\pi\times45\Hz$ and $\omega_z=2\pi\times190\Hz$. The atomic cloud, when illuminated by a
red detuned laser beam with $\Delta=-100\GHz$, is optically equivalent to a graded index lens of Gaussian profile
$~e^{-x^2/(2\sigma_z^2)-y^2/(2\sigma_y^2)-z^2/(2\sigma_z^2)}$. Its peak refractive index is $n_\text{ref}=1.0000093$ and its widths are
$\sigma_x=\sigma_y=22\text{ }\mu m$, and $\sigma_z=5.2\text{ }\mu m$.
To generate the electrostriction force we use a $\lambda=780\nm$ laser, $50-200\GHz$ detuned from the $\ket{F=2}\to\ket{F'=3}$ transition. The beam is coupled to
a polarization maintaining single mode fiber and ejects with a waist of $1.1\mm$. Under these parameters, the dipole force associated with the laser beam itself
is suppressed by $10^{-3}$ compared to the electrostriction force, and the scattering probability is only a few percent. The dipole force that the light focused
by the atoms exerts on the atoms is negligible. The electrostriction beam is shone from the $\hat{y}$ direction (see Fig.~\ref{Fig. firstfig3}). The atomic cloud
is optically extended ($\sigma\gg\lambda$), so a simple refractive media treatment is adequate. It is dilute ($nk^{-3}=0.25$), so dipole-dipole interatomic
interactions~\cite{Havey1,dalibard} do not affect our experiment. To measure the force we apply a short pulse of duration $\tp$ right after releasing the cloud,
and image the momentum distribution after a long expansion time [$18\ms$, Figs.~\ref{Fig. firstfig3}(a)-\ref{Fig. firstfig3}(c)] by absorption imaging along the
$\hat{z}$ direction. Since the force is anisotropic, the cloud expands more in the transverse directions and gains an aspect ratio (AR) larger than unity. If the
atoms do not move during the pulse (impulse approximation, $\tp\ll\omega^{-1}$) we can calculate the atomic cloud size $\sigma$ along the transverse ($\perp$) and
axial ($\parallel$) directions after time of flight. For a cloud with initial temperature $T$ and after expansion time $t$,
\begin{equation}\label{eq. EsSigmas}
\begin{split}
    &\sigma_{\perp}=\sqrt{\frac{k_B T}{m\omega_{\perp}^2}}\sqrt{\left(1-\frac{\hbar\Gamma}{k_B
    T}\frac{\Gamma}{8\Delta}\frac{I}{I_s}\omega_{\perp}^2t\tp\right)^2+\omega_{\perp}^2t^2}\\
    &\sigma_\parallel=\sqrt{\frac{k_B T}{m\omega_{\parallel}^2}}\sqrt{1+\omega_\parallel^2t^2}.
\end{split}
\end{equation}
After a long expansion time the aspect ratio $\sigma_{\perp}/\sigma_{\parallel}$ of the cloud reaches an asymptotic value,
\begin{equation}\label{eq. ARESTOF}
    \text{AR}^2=1+\left(\frac{\hbar\Gamma}{k_B T}\frac{\Gamma}{8\Delta}\frac{I}{I_s}\omega_{\perp}\tp\right)^2=1+\left(\frac{\pes}{\pth}\right)^2,
\end{equation}
where $\pes=\frac{\hbar\Gamma\sqrt{m}}{\sqrt{k_B T}}\frac{\Gamma}{8\Delta}\frac{I}{I_s}\omega_{\perp}\tp$ is the momentum distribution width of the
electrostriction impulse, and $\pth=\sqrt{mk_BT}$ - the width of the initial cloud thermal momentum distribution.

The above derivation relies on the impulse approximation. In order to check its validity, we numerically solved the dynamics of the atomic cloud when applying
electrostriction on it using a phase-space simulation. The simulation results coincide with our analytic predictions for all measurements presented here,
confirming the impulse approximation. Further theoretical considerations regarding the above derivation are detailed in~\footnote{See supplementary material,
including Refs. \cite{meystre2013elements,ritschReview,labeyrie2014optomechanical,camara2015optical,loudon2000quantum}.}.

Performing this experiment we observe that the electrostriction pulse neither changes the cloud size along the longitudinal direction nor the center of mass
[Figs.~\ref{Fig. firstfig3}(a)-\ref{Fig. firstfig3}(b)]. This indicates that our experiment suffers no significant scattering and demonstrates the transverse
nature of the optomechanical strain. This is more dramatically demonstrated performing the same measurement on a BEC. In this case [Fig.~\ref{Fig. firstfig3}(c)]
the usually fragile bimodal distribution typical of a BEC along the axial direction is unaffected by the strong momentum impulse in the transverse directions.
Similar results for pure condensates prove that the force acting on the atoms is different from that predicted in~\cite{Cattani}. We nevertheless emphasize that
our predictions in Eqs.~\eqref{eq. EsSigmas} and \eqref{eq. ARESTOF} do not hold for a BEC, for which the equation has to be modified.

Applying an electrostriction pulse \emph{in situ} generates a breathing mode oscillation, only in the transverse directions. This can be observed by letting the
cloud evolve in the trap for some variable time, and imaging it after release [Fig.~\ref{Fig. firstfig3}(e)].
The results in Fig.~\ref{Fig. firstfig3} did not depend on the laser polarization, in accordance with our theory. This observation also indicates that the
interactions we induce between atoms are not dipole-dipole interactions.

We perform strain measurements after short electrostriction pulses for a large range of detunings $|\Delta|<200\GHz$. The results (Fig.~\ref{Fig.
ARvsdetunningLogBandRfit4}) are consistent with a $1/\Delta$ rather than a $1/\Delta^2$ scaling. This agrees with our prediction in Eq.~\eqref{eq. es} and rules
out the scattering force and the forces in~\cite{kurizki2002new,Cattani}, which scale as $1/\Delta^2$, as a source of the strain observed. Imaging the cloud a
short time after the electrostriction impulse we observe the effect of the detuning's sign as well~\cite{Note1}.

\begin{figure}
    \begin{center}
        \includegraphics[width=\linewidth]{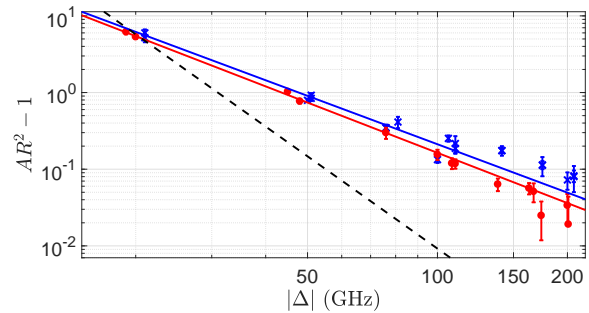}
        \caption{Scaling of strain with detuning $\Delta$. A thermal cloud AR after an electrostriction pulse and free expansion, red circles (blue crosses),
        correspond to a red (blue) detuned electrostriction laser. Fits to the data (solid lines) indicate a scaling of the force as $1/\Delta^\alpha$, with
        $\alpha=1.09(5)$ [$\alpha=1.05(9)$] for the red (blue) detuned electrostriction laser. A prediction (dashed line) based on a force that scales as
        $1/\Delta^2$ is shown as well. The error in $\alpha$ corresponds to a $95\%$ confidence level. We used a cloud with a temperature of $1.1\text{
        }\mu\text{K}$ and a laser with intensity $1.1\times10^4\mW/\cm^2$, pulsed for $0.5\ms$.}
        \label{Fig. ARvsdetunningLogBandRfit4}
    \end{center}
\end{figure}

\begin{figure}
    \begin{center}
        \includegraphics[width=\linewidth]{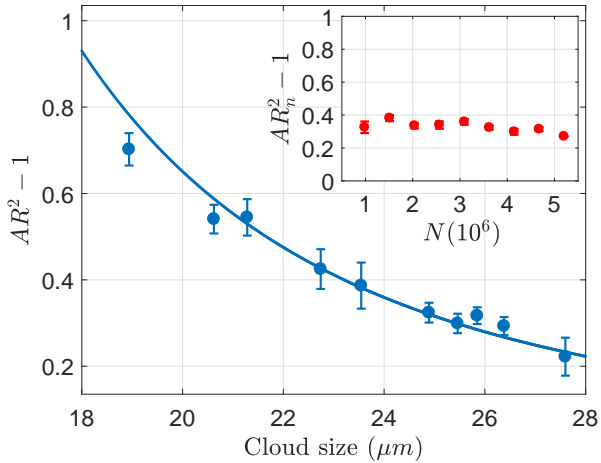}
        \caption{Strain for clouds of different sizes. Measured cloud aspect ratio after an electrostriction pulse and free expansion for different cloud sizes
        (circles), and the theoretical prediction (line, scaled strain). All data points correspond to thermal clouds besides the first one, which includes a
        small condensed fraction. Smaller clouds consist of fewer atoms, but the $\text{AR}_n$ (normalized AR~\cite{Note1}) is independent of the number of atoms
        as can be seen in the inset. We used a laser with intensity $7.4\times10^3\mW/\cm^2$ and detuning $73\GHz$, pulsed for $0.25\ms$.}
        \label{Fig. CloudSizeScan}
    \end{center}
\end{figure}

\begin{figure}
    \begin{center}
    \begin{overpic}
		[width=\linewidth]{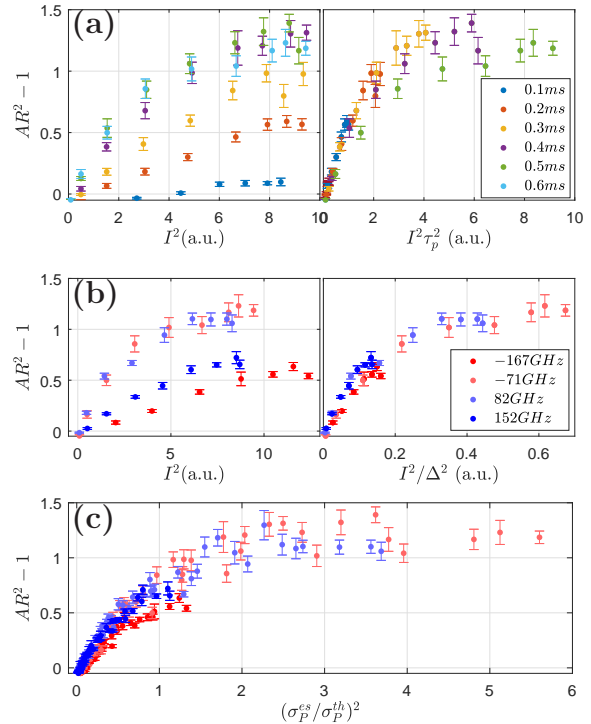}
		\put(33,275){\large \textbf{(a)}}
		\put(33,173){\large \textbf{(b)}}
		\put(33,88){\large \textbf{(c)}}
	\end{overpic}
        \caption{Strain saturation with electrostriction laser intensity $I$, detuning $\Delta$ and pulse duration $\tp$. \text{(a)} Saturation with laser
        intensity $I$ for different pulse durations $\tp$ (left graph). After scaling the results by $\tp^2$ (right) they collapse to a single curve. \text{(b)}
        Saturation with laser intensity $I$ for different detunings $\Delta$ (left graph). After scaling the results by $\Delta^2$ (right) they collapse to a
        single curve. \text{(c)} When plotted as a function of the momentum impulse $\pes$, all measurements collapse together. The laser intensity is changed
        between $0-9\times10^3\mW/\cm^2$, the detuning $\Delta$ between $(-167)-(+152)\GHz$, and the pulse duration $\tp$ between $0.1$ and $0.6\ms$. $\pth$ is
        the width of the thermal momentum distribution prior to the electrostriction impulse.}
        \label{Fig. saturation}
    \end{center}
\end{figure}

To qualitatively compare our observations to the theoretical prediction [Eq.~\eqref{eq. EsSigmas}], we carefully calibrate our experimental parameters. In
particular, we measured the spontaneous Raman transition rate between the \ket{F=1} and \ket{F=2} hyperfine states due to the electrostriction laser. The measured
rate was in accordance with the rate calculated~\cite{Note1} using the Kramers-Heizenberg equation~\cite{loudon2000quantum,cline1994spin}, given the independently
directly measured laser intensity and detuning values, and the atomic parameters~\cite{steck2016rubidium}. After calibration, the observed effect is roughly $2.5$
times weaker than expected. As we currently do not have an explanation for this discrepancy, we scale our predictions by this factor when comparing results to
theory throughout this paper (Figs.~\ref{Fig. ARvsdetunningLogBandRfit4}-\ref{Fig. saturation}).

We further investigated the dependence of the electrostriction force on the cloud parameters: total number of atoms $N$ and cloud size. We measured the aspect
ratio, $N$, and the cloud size, while applying the same strain pulse on the cloud (Fig.~\ref{Fig. CloudSizeScan} and inset). As seen, the measured AR is
independent of $N$, as expected from Eq.~\eqref{eq. es}. On the other hand, the effect shows a strong dependence on the atomic cloud size. Decreasing the cloud
size makes the cloud a stronger lens, causing the beam to focus stronger and impart more momentum on the atoms.

The dipole force might, in principle, cause dependence on the cloud size if the laser beam deviates from a plane wave, suffering intensity profile changes on
length scales comparable with the cloud size. In order to avoid such situations, we work with a beam size about 100 times greater than our cloud size. We avoid
speckles using a single mode fiber with a collimator and no other optical elements before the vacuum cell. We verified the absence of spatial sharp intensity
changes by direct imaging of the beam. The strain we observed did not change after a slight misalignment of the beam, suggesting that indeed no significant local
gradients appear. This shows that the observed cloud size dependence is not due to a dipole force of the electrostriction beam.

In order to verify the linearity of the electrostriction force strength with intensity $I$, we measured the strain as a function of growing optical power and
different pulse durations and detunings. As seen in Figs.~\ref{Fig. saturation}(a) and \ref{Fig. saturation}(b), linearity is indeed evident for low intensities.
However, a clear saturation of the strain [Figs.~\ref{Fig. saturation}(a)-\ref{Fig. saturation}(c)] occurs at high intensities, for various electrostriction pulse
durations and detunings.
We measured the dependence of saturation on the cloud temperature as well (not shown in~Fig.~\ref{Fig. saturation}), and found it appears to depend on the impulse
applied to the atomic cloud $I\tp/(T\Delta)$. This is evident from the collapse of all data on a single curve as in~Fig.~\ref{Fig. saturation}. We note that the
results presented in Figs~\ref{Fig. ARvsdetunningLogBandRfit4} and~\ref{Fig. CloudSizeScan} were performed for unsaturated strain.

The saturation of the effect stems neither from changes in the internal state of the atoms nor from expansion of the cloud during the pulse. Our pulses are
considerably short (up to $1\ms$) compared with the trap oscillation period of typically $20\ms$ and the scattering rate of $20\Hz$. We verified that there are no
changes in the cloud density and internal state by imaging the cloud at short times and measuring the number of atoms in the \ket{F=1} hyperfine state. The only
evident change is the momentum distribution of the atoms, which should not affect the strain via our theory. As is clear from Fig.~\ref{Fig. saturation}(c),
saturation occurs when the atoms have accelerated to a momentum roughly equal to their initial thermal velocity spread, $\pes=\pth$.

The observation that lensing saturates close to $\pes=\pth$ is reminiscent of a classical version of Einstein's recoiling-slit
gedankenexperiment~\cite{schilpp1951albert,jammer1974philosophy}. In this experiment an interference pattern of light that passed scatterers (slits) is dephased
when the momentum imparted to the scatterers by the photons separates the scatterers in momentum space giving away the which-path information. In our experiment,
lensing occurs due to coherent interference of light passing through different parts of the cloud. In an analogy to the above gedankenexperiment, the cloud would
therefore cease to behave as a coherent lens after accumulating a momentum impulse $\pes$ comparable to their initial momentum distribution $\pth$.

The bound on the electrostriction momentum given to the atomic cloud may prevent application of electrostriction for long times. For short times, the
optomechanical strain has some interesting features of potentially practical importance (details in~\cite{Note1}). An electrostriction laser beam applied to a BEC
can effectively modify the interparticle interaction strength at the mean-field level, mimicking the effect of a Feshbach resonance, without really changing the
scattering length. Interaction tuning was used before~\cite{cetina2016ultrafast} for short times using an optical Feshbach resonance. A BEC with attractive
effective interactions induced by an electrostriction laser is unstable to spatial density modulations seeded by initial noise in the density profile of the
cloud, as in nonlinear optical fibers~\cite{ModulationalInstab1986}. An atomic cloud with repulsive effective interactions works to smoothen out spatial density
modulations. This can serve as an explanation to the unexplained red-blue asymmetry in~\cite{labeyrie2014optomechanical}. The electrostriction potential
[Eq.~\eqref{eq. es}] serves as a logarithmic nonlinearity, and thus a BEC under illumination can support stable solitons in any dimension~\cite{LogNonLin} - a
nontrivial feature~\cite{pedri2005two,fleischer2003observation}. Finally, a thermal atomic cloud can be self-trapped by its own strain, resembling a bright
soliton~\cite{khaykovich2002formation} in the transverse directions, incoherent and with arbitrary shape and size.

In summary, we report the observation of optomechanical strain applied to \Rb thermal and condensed atoms when illuminated by an intense, far detuned homogeneous
laser beam. We experimentally demonstrate the basic features of electrostriction, distinguishing it from the well-established scattering and dipole forces, and
proving that it is a new type of force acting on cold atoms. By the observed electrostriction characteristics, we point out that this force is distinct from
theoretically predicted light-induced forces such as those discussed in~\cite{mazets2000ground,kurizki2002new,Cattani} or collective forces measured
in~\cite{meir2014cooperative,pellegrino2014observation}. The experimental results are in qualitative agreement with our theory. Electrostriction has the potential
to be an important tool in cold atom experiments as it effectively induces interparticle interactions, which can be optically tuned.

\begin{acknowledgements}
The authors thank Igor Mazets, Gad Afek and Arnaud Courvoisier for discussions. This work was supported by the Crown photonics center, ICore Israeli excellence
center Circle of light, and the European research council (Consolidator Grant No. 616919). N.M. and H.E. contributed equally to this work.
\end{acknowledgements}

\bibliographystyle{apsrev4-1}
\bibliography{ESbib}

\end{document}


\heading{Supplementary Material for Observation of Optomechanical Strain in a Cold Atomic Cloud}
\begin{center} Noam Matzliah, Hagai Edri, Asif Sinay, Roee Ozeri and Nir Davidson\end{center}
\begin{center} Department of Physics of Complex Systems, Weizmann Institute of Science, Rehovot 7610001 , Israel\end{center}

\subsection*{Derivation of the electrostriction force and its effect in time of flight measurements}
In this section we derive an expression (Eq.(1)) for the electrostriction force under research in this work, and its effect on time of flight measurements (Eq.(2)). These derivations are original.

A plane wave propagating in the $\hat{z}$ direction has a phase of $(k_Lz-\omega_L t)$. $k_L$ denotes the wave number of the incident light, and $\omega_L$ - its angular frequency.

After passing through an infinitesimal section of width $dz$ in an atomic cloud, the light will acquire a phase of $\phi=k_L n_{ref}dz$, where the local refractive index of the cloud is given by $n_{eff}(\vec{r})=1+\Re e(\tilde{\chi})/2$. Here $\tilde{\chi}$ is the Fourier transform of the atoms electric susceptibility. An expression for $\tilde{\chi}$ is derived by solving the optical Bloch equations for two-level atoms~\cite{meystre2013elements}

\begin{equation}\label{eq. suscept}\tag{S1}
    \tilde{\chi}=i\frac{3}{8\pi^2}n\lambda^3(\rho_{11}-\rho_{22})\frac{\Gamma}{\frac{\Gamma}{2}-i\Delta}.
\end{equation}
Here $n$ is the atomic cloud density, $\lambda$, the light wavelength, $\rho_{11}$ and $\rho_{22}$, the populations (the diagonal elements of the density matrix $\rho$) of the two atomic states denoted $\ket{1}$ and $\ket{2}$, $\Gamma$, the width of the atomic transition, and $\Delta$, the detuning of the light.

In our experiment we use a far detuned laser $|\Delta|\gg\Gamma$ and to a good approximation the atoms stay in the ground state so that $\rho_{11}\approx1$ and $\rho_{22}\approx0$. Under these conditions we can approximate the phase $\phi$ by
\begin{equation}\label{eq. AtomsPhase}\tag{S2}
    \phi\approx k_L dz-\frac{\sigma_0}{4}\frac{\Gamma}{\Delta}ndz.
\end{equation}
$\sigma_0=3\lambda^2/(2\pi)$ being the cross section for photon scattering from an atom. After passing the infinitesimal section at point $(x_0,y_0,z_0)$, the light will have a phase of $(\phi(x,y,z_0)+k_L z-\omega_L t)$ at point $(x,y,z)$. The corresponding Poynting vector under the eikonal approximation~\cite{born1980principles} takes the form $\vec{S}\approx\omega_L^{-1}|\vec{E}|^2\vec{\nabla}(\phi(x,y,z_0)+k_L z)/(2\mu_0)=\omega_L^{-1}|\vec{E}|^2(k_L\hat{z}+\vec{\nabla}_{x,y}\phi)/(2\mu_0)$, where $\mu_0$ is the vacuum permeability and $\vec{E}$, the electric field amplitude associated with the light. The Poynting vector is the electromagnetic energy $E_\gamma$ flux. Using the dispersion relation of light $E_\gamma=cP_\gamma$, the electromagnetic momentum $\vec{P}_\gamma$ flux is $\vec{S}/c$. By momentum conservation, the momentum change per unit time in an infinitesimal section is $d\vec{P}_a/dt=\oiint dA_{\perp}\vec{S}/c$, where the integration is over surface elements $dA_{\perp}=d\vec{A}\cdot\hat{S}$ surrounding the section. This momentum change corresponds to the total force on the atoms in the section $\vec{f}_{total}=d\vec{P}_a/dt$. Dealing with clouds which cause only slight changes in the Poynting vector, its direction will stay approximately the same, and thus $\frac{1}{c}\oiint dA_{\perp}\vec{S}\approx\frac{1}{c}\Delta\vec{S}dA$, where $\Delta\vec{S}$ is the difference in the Poynting vector after and before passing the infinitesimal section, and $dA$ is the area of incidence of the section. Plugging the speed of light $c=\frac{1}{\sqrt{\varepsilon_0\mu_0}}$, the electric field intensity $I=c\varepsilon_0|\vec{E}|^2/2$ and the vacuum permittivity $\varepsilon_0$, we get
\begin{equation}\tag{S3}
    \vec{f}_{total}=-\frac{k}{2\mu_0 \omega_L^2}|\vec{E}|^2\vec{\nabla}_{x,y}\phi dA=\frac{\sigma_0}{4}\frac{\Gamma}{\Delta}\frac{I}{\omega}\vec{\nabla}_{x,y}ndzdA.
\end{equation}
The total force is equally distributed among $dN=ndzdA$ atoms consisting the infinitesimal section, and thus each atom feels a force of
\begin{equation}\label{eq. ESsup}\tag{S4}
    \vec{f}_{es}=\frac{\vec{f}_{total}}{dN}=\frac{\sigma_0}{4}\frac{\Gamma}{\Delta}\frac{I}{\omega_L}\frac{\vec{\nabla}_{x,y}n}{n}=\frac{\hbar\Gamma^2}{8\Delta}\frac{I}{I_\text{s}}\frac{\vec{\nabla}_{\perp}n}{n},
\end{equation}
where we used the relation between the scattering cross section and the saturation intensity $\sigma_0=\hbar\omega_L\Gamma/(2I_s)$.

We note that the cloud imprints a phase on the laser light passing through it, and this phase does not translate to intensity gradients by means of free propagation in the cloud itself. This is in contrast to cavity mediated interaction schemes described in~\cite{ritschReview}, and to half-cavity schemes as in~\cite{labeyrie2014optomechanical,camara2015optical}. In these works, the round trip in the cavity translates laser phase changes into intensity changes felt by the atoms.

For a nondegenerate cloud with temperature $T$, the phase space distribution of the atomic cloud in a harmonic trap is $f_0(\vec{r},\vec{p})\approx f_0 e^{\beta(\mu-H)}$. Here $H=p^2/(2m)+U(\vec{r})$ is the single particle Hamiltonian, $U=m/2\sum_{i=1}^3\omega_i^2x_i^2$, the trapping potential, $\beta=1/(k_B T)$, $m$, the atomic mass, $\vec{r}$ and $\vec{p}$, position and momentum, $\omega_i$, the harmonic trap angular frequencies, and $f_0$, a normalization constant setting the integral $1/h^3\int d^3pd^3r f_0(\vec{r},\vec{p})=N$ to the total number of atoms in the cloud $N$. The spatial density distribution of the cloud in the trap is $n_0(\vec{r})=1/h^3\int d^3p f_0(\vec{r},\vec{p})=f_0/(\lambda_{th}^3)e^{\beta(\mu-H)}$, where $\lambda_{th}=\sqrt{2\pi\hbar^2/(mk_BT)}$ is the De-Broglie thermal wavelength of the atoms. An electrostriction beam will apply to the cloud a force \eqref{eq. ESsup} of the form
\begin{equation}\tag{S5}
    \vec{f}_{es}=m\beta\frac{\hbar\Gamma^2}{4\Delta}\frac{I}{I_s}\sum_{i=x,y,z}\omega_i^2x_i\hat{x}_i.
\end{equation}
After shining the laser beam along $\hat{z}$ for a short time $\tau_p\ll\omega_i^{-1}$, the phase space distribution of the atoms will be $f_1(\vec{r},\vec{p})=f_0(\vec{r},\vec{p}-\tau_p\vec{f}_{es}(\vec{r}))$. If the cloud is released from its trap right after the laser pulse and expends ballistically, its density distribution will be:
\begin{equation}\tag{S6}
\begin{split}
    &n_{TOF}(\vec{r},t,\tau_p)=\frac{1}{h^3}\int d^3p f_1(\vec{r}-\frac{\vec{p}}{m}t,\vec{p})=\left(\prod_{i=x,y,z} b_i\right)^{-1} n_o(\frac{x}{b_x},\frac{y}{b_y},\frac{z}{b_z})\\
    &b_{x,y}(t,\tau_p)=\sqrt{(1-\tau_p\beta\frac{\hbar\Gamma^2}{8\Delta}\frac{I}{I_s}\omega_{x,y}^2t)^2+\omega_{x,y}^2t^2}\\
    &b_z(t)=\sqrt{1+\omega_z^2t^2}.
\end{split}
\end{equation}
$b_i$ are the expansion factors of the cloud during ballistic expansion along $\hat{x}_i$. The sizes of the cloud during time of flight are thus given by
\begin{equation}\label{eq. EsSigmas}\tag{S7}
\begin{split}
    &\sigma_{x,y}=\sqrt{\frac{k_\text{B} T}{m\omega_{x,y}^2}}\sqrt{\left(1-\frac{\hbar\Gamma}{k_\text{B} T}\frac{\Gamma}{8\Delta}\frac{I}{I_\text{s}}\omega_{x,y}^2t\tp\right)^2+\omega_{x,y}^2t^2}\\
    &\sigma_z=\sqrt{\frac{k_\text{B} T}{m\omega_z^2}}\sqrt{1+\omega_z^2t^2}.
\end{split}
\end{equation}

\subsection*{Experimental conditions and analysis specifics}
The experimental parameters used in our measurements presented in the letter are summarized in Table~\ref{Tab. ExpParams}.
\begin{table}[h]
\caption{Experimental parameters used in our measurements.}
\label{Tab. ExpParams}\centering
\begin{tabular}{ccccc}
\hline
Parameter&Fig .1(b)&Fig .2&Fig .3&Fig .4\\
\hline
\hline
$N(10^6)$&$1.2$&$3$&$1.5$ - $4.5$&$2$\\
$T(\mu K)$&$0.4$&$1.1$&$0.26$ - $1.1$&$0.5$\\
$I\left(\frac{\mW}{\cm^2}\right)$&$8\times10^3$&$1.1\times10^4$&$7.4\times10^3$&$<9\times10^3$\\
$\Delta(\GHz)$&$47$&$-200$ - $200$&$73$&$-167$ - $152$\\
$\tp(\ms)$&$0.5$&$0.2$&$0.25$&$<0.6$\\
$\frac{\omega_{\text{trap}}}{2\pi}(\Hz)$&$45\times45\times190$&$64\times64\times270$&$57\times57\times240$&$49\times49\times208$
\end{tabular}
\end{table}
All measurements were performed waiting a time of flight of about $18\text{ }\ms$. We note that pulsing the cloud \emph{in situ} right before release or right after release does not affect the result.

While repeating the measurements in Fig. 2 for each detuning, the electrostriction laser intensity $I$ and the cloud temperature $T$ slightly fluctuate by a few percent. We monitor $I$ by a photodiode and $T$ by the clouds width along the longitudinal direction for each run. In Fig. 2 we thus plot the normalized $\text{AR}^2-1$ as the combination $(\text{AR}^2-1)\left(\frac{<I>}{I}\frac{T}{<T>}\right)^2$ in the ordinate, where $<I>$ and $<T>$ are the average laser intensity and cloud temperature respectively. This slightly affected the fit parameter $\alpha$ (see Fig. 2) by only a few percent.

Although clearly excluding the scattering force as responsible to the observed strain, the results in Fig. 2 slightly disagree with our theoretical prediction. We attribute this to systematic errors, presumably resulting from the long total time needed for data collection, and the fact the detuning is tuned manually.

Regarding Fig. 3, changing the number of atoms $N$, while keeping a constant cloud size (or equivalently temperature $T$) is not straightforward, since the evaporative cooling of the atoms is affected by the number of atoms involved via the collision rate. We thus varied both $N$ and $T$ and plotted (Fig. 3 inset) the normalized $\text{AR}_n^2-1$ as the combination $(\text{AR}^2-1)\left(\frac{T}{T_0}\right)^2$ in the ordinate, where $T_0=540\text{ }\nK$ is a typical cloud temperature corresponding to a cloud size of $20\text{ }\mu m$. We omitted all data points in the inset having less than $3\times10^5$ atoms, which suffer larger errors due to low signal to noise ratio. We omitted all data points in the inset having temperatures less than $0.8\text{ }\mu K$ in order to avoid condensates. The data points were clustered into ten bins.

A typical scattering rate for our electrostriction laser  is $80\text{ }\Hz$ for a laser of power $150\text{ }\mW$ (peak intensity $8\times10^3\text{ }\frac{\mW}{\cm^2}$) and $100\text{ }\GHz$ detuning.  For a pulse time of $0.2\text{ }\ms$, only $2\%$ of the atoms will recoil on average. The ratio between the dipole potential of the electrostriction laser,
\begin{equation}\tag{S8}
\begin{split}
    &U_\text{dipole}^{es}=-\frac{m}{2}\omega_\text{dipole}^2x^2\\
    &\omega_\text{dipole}^2=\frac{2}{mw_0^2}\frac{\hbar\Gamma^2}{4\Delta}\frac{I_0}{I_s}
\end{split}
\end{equation}
and the electrostriction potential,
\begin{equation}\tag{S9}
\begin{split}
    &U_\text{es}=-\frac{\hbar\Gamma^2}{8\Delta}\frac{I_0}{I_\text{s}}\ln\left(\frac{n}{n_0}\right)\\
    &n=n_0e^{-\frac{U_\text{ext}}{k_BT}}\\
    &U_\text{ext}=-\frac{m}{2}\omega_\text{x}^2x^2
\end{split}
\end{equation}
along $\hat{x}$ is
\begin{equation}\tag{S10}
    \frac{U_\text{dipole}^{es}}{U_\text{es}}=\frac{2k_BT}{\frac{m}{2}\omega_\text{x}^2w_0^2}.
\end{equation}
Here $m$ is the atoms mass, $w_0=1.1\text{ }\mm$, the electrostriction laser beam waist, $I_0$, the laser's peak intensity, and $\omega_{x}=2\pi\times45\text{ }\Hz$, the trap angular frequency along $\hat{x}$.

For a cloud of temperature $T=400\text{ }\nK$, we get $U_\text{dipole}/U_\text{es}=1.6\times10^{-3}$. These calculations demonstrate the dipole force of the electrostriction laser and scattering from it are avoidable in our experiment.

Having a nonisotropic Gaussian density profile, the cloud behaves a lens with aberrations and astigmatism. The paraxial focal length of the cloud along $\hat{x}$ is $f_x=934\text{ }\mm$. The distance from the cloud where the Fresnel number becomes unity along $\hat{x}$ is $L_x=0.6\text{ }\mm$. The cloud thus behaves more as a diffractive element than a refractive one. Notice that the length scale for changes in the electrostriction beam intensity profile is $L_x/\sigma_x=28$ times larger than the cloud size. This illustrates that the cloud fulfills the "thin lens approximation" allowing us to optically analyze it considering only phase imprinting. As another result from this, the dipole force acting on the cloud, resulting from the intensity gradient of the laser beam after lensing in the cloud, is suppressed - $10^{-2}$ times weaker than the measured electrostriction force in a typical case. The dipole force due to lensing scales as $1/\Delta^2$, while the electrostriction force scales as $1/\Delta$.

\subsection*{The sign of detuning effect on the strain}
\begin{figure}
    \begin{center}
        \includegraphics[width=\linewidth]{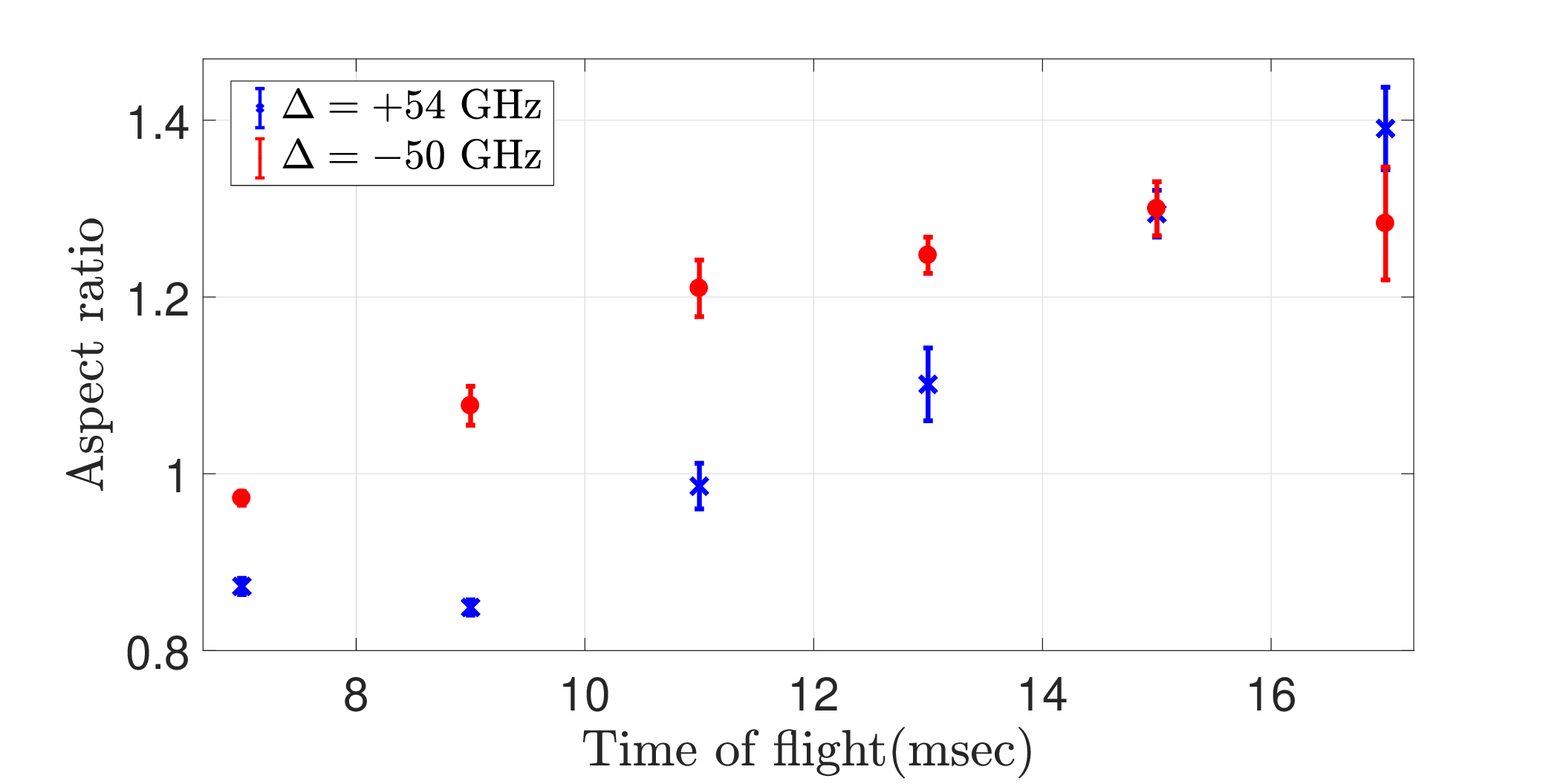}
        \caption{Comparison between the atomic cloud aspect ratio evolution for a red (red dots) and a blue (blue crosses) detuned laser beam pulse. The cloud is released from the trap, allowed to freely expand for $6\ms$, pulsed during $1\ms$, continue expanding freely for some time of flight, and imaged. A red detuned beam makes the cloud to expand in the transverse direction. A blue detuned beam makes the cloud to shrink in the transverse direction and then expand.}
        \label{Fig. BvsR}
    \end{center}
\end{figure}

The method presented in the manuscript is inert to the electrostriction laser beam detuning sign for long time of flight (see Eq.(3)). In principle, imaging the atoms after a short time of flight can reveal the difference between a blue and a red detuned electrostriction laser beam. For a typical working point one needs to image the cloud after about $1\text{ }\ms$ suffering a trade-off between the strength of the effect and the time window for observing it. This also poses a challenge to our imaging system due to the large optical density of the cloud, and limited resolution. We thus modified the method releasing the cloud from the trap, allowing it to freely expand for $6\text{ }\ms$, pulsed it during $1\text{ }\ms$, let it continue expanding freely for some variable time of flight, and imaged it - see Fig.~\ref{Fig. BvsR}. This way the cloud is already large when imaged and the effect lasts for a few $\ms$.
A red detuned laser pulse exerts a momentum kick outwards, making the cloud expand in the transverse direction and its aspect ratio to exceed unity. A blue detuned laser pulse exerts a momentum kick inwards, making the cloud shrink in the transverse direction and its aspect ratio to drop below unity. After the cloud shrinks, it expands again and the aspect ratio exceeds unity.

This demonstrates the effect of the detuning's sign, though comparison to simulation shows the cloud's aspect ratio should have dropped much lower for blue detuning. This relates to our observation of saturation in which after gaining a momentum impulse comparable to that of the thermal momentum distribution width, the cloud ceases to further gain momentum. In the measurement here the cloud after time of flight ceases to gain considerable momentum. In both cases, induced position-momentum correlations in phase space prevents additional momentum transfer. In the first case, position-momentum correlations are induces by electrostriction, while in the second case it is induced by free expansion.

\subsection*{Multilevel atom treatment}
The theoretical treatment presented in this work considered the susceptibility of two-level atoms. The susceptibility of \Rb atoms take a different value, which depends on the laser polarization.

For a multi-level atom, the susceptibility takes the form
\begin{equation}\tag{S11}
    \chi=n\frac{\frac{i}{\hbar}|\mu|^2}{\epsilon_0(i\Delta+\Gamma)}.
\end{equation}
Scattering atoms from $\ket{F=1,m_F=1}$, the electric dipole moment takes the form
\begin{equation}\tag{S12}\label{eq. musqr}
    |\mu|^2=\Sigma_{F',m_F'}\left|\Sigma_{q}a_q\bra{F=1,m_F=1}er_q\ket{F',m_F'}\right|^2,
\end{equation}
where $q$ is the circular decomposition index, $a_q$ is defined by the decomposition $\vec{r}\cdot\hat{\epsilon}=\Sigma_{q\in\{0,\pm1\}}a_qr_q$, $\hat{\epsilon}$ being the laser polarization unit vector. Note that the sum over $m_F$ and $q$ reduces to a single sum due to selection rules. The polarization of a general elliptically polarized laser propagating along $\hat{y}$ toward atoms quantized along $\hat{z}$, as in our experiment, can be parameterized as $\hat{\epsilon}=\cos{\phi}\hat{z}+\sin{\phi}e^{i\theta}\hat{x}$. We thus recognize $a_0=\cos{\phi}$ and $a_{\pm1}=\frac{\sin{\phi}e^{i\theta}}{\sqrt{2}}$. Plugging all matrix elements~\cite{steck2016rubidium} in Eq.\eqref{eq. musqr} in terms of the reduced matrix element (RME) gives,
\begin{equation}\tag{S13}\label{eq. mu1}
\begin{split}
    &|\mu|^2=\frac{1}{3}\text{RME}^2\\
    &\text{RME}=\bra{J=\frac{1}{2}}||er||\ket{J'=\frac{3}{2}}=3.584\times10^{-29}\text{ }\text{C}\cdot\text{m}.
\end{split}
\end{equation}
Notice the susceptibility is independent of the laser polarization. Furthermore, it takes the same value as for the cycling transition for a linearly polarized laser
\begin{equation}\tag{S14}
    |\mu^{\pi\text{-pol}}_{\text{cycling transition}}|^2=\frac{1}{3}\text{RME}^2.
\end{equation}
We note this simple result is valid for elliptically polarized laser propagating perpendicular to the quantization axis. For instance, a laser propagating parallel to the quantization axis will induce a polarization-dependent effect.

This proves we can treat our atoms as two-level systems using the cross section~$\sigma_0=1.938\text{ }\times10^{-9} \cm^2$, or equivalently the saturation intensity~$I_\text{s}=2.503\text{ }\frac{\mW}{\cm^2}$, of a linearly polarized laser working on the cycling transition.

\subsection*{Spontaneous Raman transition rate}
We measured a spontaneous Raman transition rate of $\Gamma^\text{exp}_{\ket{i}\to\ket{f}}=15.4\text{ }\Hz$ between the $\ket{i}=\ket{F=1,m_F=1}$ to any $\ket{f}=\ket{F=2,m_F}$ hyperfine state due to the $\pi$-polarized electrostriction laser of intensity $I=8.4\times10^3\text{ }\frac{\mW}{\cm^2}$ (total power of $P=160\text{ }\mW$) red detuned by $\Delta=2\pi\times100\text{ }\GHz$ from the $\ket{F=2}\to\ket{F'=3}$ transition.

The rate of photon scattering events in which an atom initially in state $\ket{i}$ ends up in state $\ket{f}$ is given by the Kramers-Heisenberg formula~\cite{loudon2000quantum,cline1994spin},
\begin{equation}\tag{S15}\label{eq. KramersHeisenberg}
    \Gamma_{i\to f}=g^2\Gamma\left|\frac{a_{i\to f}^{(1/2)}}{\Delta}+\frac{a_{i\to f}^{(3/2)}}{\Delta-\Delta_f}\right|^2.
\end{equation}
Here, $g=\frac{E\mu}{2\hbar}$, $E=\sqrt{\frac{2I}{c\epsilon_0}}$ is the laser-beam electric field amplitude, $c$, the speed of light, $\epsilon_0$, the vacuum dielectric constant, and\\$\mu=|\bra{2P_{3/2},F=3,m_F=3}\vec{d}\cdot\hat{\vec{\sigma}}_+\ket{2S_{1/2},F=2,m_F=2}|$, where $\vec{d}$ is the electric dipole operator. The effective amplitude\\$a_{i\to f}^{(J)}=\Sigma_q\Sigma_{e\in J}\bra{f}\vec{d}\cdot\hat{\vec{\sigma}}_q\ket{e}\bra{e}\vec{d}\cdot\hat{\vec{\sigma}}_k\ket{i}/\mu^2$ is the sum over amplitudes of scattering through all levels, $\ket{e}$, in the $^2P_J$ manifold, $\Delta$ is the laser detuning from the $^2S_{1/2}\to^2P_{1/2}$ transition, and $\Gamma=2\pi\times6.0666\text{ }\MHz$ is the radiative linewidth of the excited states in the $^2P$ manifold~\cite{steck2016rubidium}.

In our case, we use detuning $\Delta$ up to few hundreds of $\GHz$ from the $\ket{F=2}\to\ket{F'=3}$ transition. We can thus neglect the $J=3/2$ term in Eq.~\eqref{eq. KramersHeisenberg} for all working points. The matrix element $\mu$ can be written in terms of the reduced matrix element (RME - see Eq.~\eqref{eq. mu1} above),
\begin{equation}\tag{S16}\label{eq. mu}
    \mu=\bra{F'=3,m'_F=3}\vec{d}\cdot\hat{\vec{\sigma}}_+\ket{F=2,m_F=2}=\sqrt{\frac{1}{2}}\text{RME}.
\end{equation}
In our case of a $\pi$-polarized laser,
\begin{equation}\tag{S17}
\begin{split}
    \Sigma_{mF}&\left|\Sigma_e\ket{f}\vec{d}\cdot\hat{\vec{\sigma}}_q\ket{e}\bra{e}\vec{d}\cdot\hat{\vec{\sigma}}_k\ket{i}\right|^2=|\text{RME}|^4\cdot\text{CG}\\
    \text{CG}=&\left|-\sqrt{\frac{1}{8}}\cdot-\sqrt{\frac{1}{12}}+\sqrt{\frac{5}{24}}\cdot\sqrt{\frac{1}{20}}\right|^2+\left|-\sqrt{\frac{1}{8}}\cdot\sqrt{\frac{1}{24}}+\sqrt{\frac{5}{24}}\cdot\sqrt{\frac{1}{40}}\right|^2+\\
    &\left|-\sqrt{\frac{1}{8}}\cdot\sqrt{\frac{1}{8}}+\sqrt{\frac{5}{24}}\cdot\sqrt{\frac{1}{120}}\right|^2.\\
\end{split}
\end{equation}
All Clebsch-Gordan coefficients were taken from~\cite{steck2016rubidium}. Plugging all values in Eq.~\eqref{eq. KramersHeisenberg} we obtain $\Gamma^\text{theory}_{\ket{i}\to\ket{f}}=17.2\text{ }\Hz$ agreeing with the experimental value up to $10\%$. This result combined with our independent measurements of all laser and atoms parameters, convinces us we have control over the experiment parameters.

\subsection*{Self trapping}
With a blue detuning, the strain laser can be adjusted to cause  a  thermal  atomic  cloud  to be  self-trapped  by  its own strain. A thermal cloud trapped in some external potential $U_{\text{ext}}(\vec{r})$ will have a Maxwell-Boltzmann spatial density $n(\vec{r})\propto e^{-\beta U_{\text{ext}}}$. Under the effect of an electrostriction laser it will experience a force (Eq.(1))
\begin{equation}\tag{S18}\label{eq. ST}
    \vec{f}_{\text{es}}\propto\frac{\vec{\nabla}_{\perp}n}{n}\propto\vec{\nabla}_{\perp}U_{\text{ext}}
\end{equation}
proportional to the force applied by $U_{ext}(\vec{r})$ in the directions transverse to the electrostriction laser beam. By turning off the external potential and rapidly turning on a blue detuned laser one can demonstrate self-trapping of the cloud in the transverse directions. This will be achieved choosing a working point at which the relation
\begin{equation}\tag{S19}\label{eq. STcond}
    \frac{\hbar\Gamma}{k_B T}\frac{\Gamma}{8\Delta}\frac{I}{I_\text{s}}=1
\end{equation}
holds. This relation can be fulfilled, with a laser of power $P=220\text{ }\mW$ and detuning $2\text{ }\THz$, suffering a scattering rate of only $0.3\text{ }\Hz$. Notice this exotic effect is predicted to work for external potentials of any shape or origin. A more involved scheme using two laser beams can be applied to get self-trapping in all three dimensions.

In steady state such considerations would imply the electrostriction force, which optical dipole trap beams exert on the trapped atoms, is comparable to the trapping force itself. If this was true, the breathing and dipole modes of noninteracting thermal atoms in a dipole trap should deviate considerably from two. No such deviation is observed, probably due to the saturation we measured at long times and since Eq.(1) ignores light momentum redistribution associated with the trapping mechanism itself. The validity of \eqref{eq. ST} and \eqref{eq. STcond} for steady state and for inhomogeneous beams is questionable.

\subsection*{Effective interaction tuning}
An electrostriction laser beam applied to a two-dimensional BEC with a homogeneous density $|\psi(\vec{r})|^2=n(\vec{r})=n_\text{0}$ will exert a potential (Eq. (1)) which can be expanded as,
\begin{equation}\label{eq. UesLin}\tag{S20}
    U_\text{es}\approx-\frac{\hbar\Gamma^2}{8\Delta}\frac{I}{I_\text{s}}\frac{n(\vec{r})-n_\text{0}}{n_\text{0}}
\end{equation}
for small deformations of the spatial BEC density profile. Plugging Eq.\eqref{eq. UesLin} in the Gross-Pitaevski equation governing the BEC dynamics we get:
\begin{equation}\label{eq. GPE}\tag{S21}
\begin{split}
    i\hbar\frac{\partial\psi}{\partial t}&=\left(-\frac{\hbar^2}{2m}\nabla^2+g|\psi|^2+U_\text{es}\right)\psi\\
    &\approx\left(-\frac{\hbar^2}{2m}\nabla^2+\tilde{g}|\psi|^2+const\right)\psi\\
    \tilde{g}=g&-\frac{\hbar\Gamma^2}{8\Delta}\frac{I}{I_\text{s}}\frac{1}{n_\text{0}}.
\end{split}
\end{equation}
The electrostriction laser effectively modifies the interparticle interaction strength $g$ at the mean-field level, mimicking the effect of a Feshbach resonance, without really changing the scattering length. The interaction can be made repulsive (attractive) using a red (blue) detuned laser. A laser of power $P=4$~W and detuning $4\text{ }\THz$ can effectively modify the scattering length of \Rb to be about $60$ times larger, suffering a scattering rate of only $1.4$~Hz. One can extend the scheme we suggest to a three-dimensional BEC using two perpendicular electrostriction beams, avoiding interference by means of a frequency shift and perpendicular polarizations.

\subsection*{Pattern formation}
A BEC with attractive effective interactions induced by an electrostriction laser will be unstable to spatial density modulations, seeded by some noise in the cloud density profile. The pattern formation process within a BEC can be derived using the Gross-Pitaevski equation (Eq.~(5)), which is identical in form to the wave equation describing light propagation in an optical fiber with a Kerr nonlinearity~\cite{ModulationalInstab1986}. We can thus rewrite the result in this reference for our case obtaining the dispersion relation $\omega(k)$ for disturbances of the BEC over its unstable stationary state
\begin{equation}\tag{S22}
    \omega^2=\left(\frac{\hbar k^2}{2m}\right)^2\left(1+\frac{2n_0\tilde{g}}{\frac{\hbar k^2}{2m}}\right),
\end{equation}
where $\tilde{g}$ was defined in Eq.~(5). One can see that when the electrostriction-induced effective attraction overcomes the repulsive background interaction ($\tilde{g}<0$), the angular frequency $\omega$ becomes imaginary for $0<k<k_\text{c}$, where
\begin{equation}\tag{S23}
    \frac{\hbar k_\text{c}^2}{2m}=-2n_0\tilde{g}.
\end{equation}
In this regime, the angular frequency gets a maximal amplitude value at $k=k_\text{p}$, where $k_\text{p}^2=k_\text{c}^2/2$.

A modulation having a wave number $k_\text{p}$ stemming from a BEC density fluctuation or an electrostriction beam intensity fluctuation, will grow exponentially faster than in any other wave number. The BEC density profile will thus get increasingly modulated at wave number $k_\text{p}$ and will reach a point, where the small deformation approximation used in deriving Eq.~(5) breaks down. We expect an eventual stabilization of the process, since the natural repulsion of \Rb atoms, which is linear in the density, will overcome the electrostriction potential, which is logarithmic in the density. This analysis is similar to the one in~\cite{Cattani} in the context of cold atoms.

\pagebreak
\bibliographystyle{apsrev}
\bibliography{ESbibSI}